\documentclass[runningheads]{llncs}

\usepackage[left=3.4cm,right=3.4cm,marginparwidth=1in]{geometry}

\usepackage[T1]{fontenc}

\usepackage{graphicx, hyperref, color, makecell}

\usepackage{setspace}

\begin{document}

\title{A Platform-Agnostic Multimodal Digital Human Modelling Framework: Neurophysiological Sensing in Game-Based Interaction}
\titlerunning{A Platform-Agnostic Multimodal DHM Framework}

\author{Daniel J. Buxton\inst{1}\orcidID{0000-0002-8729-3736} \and
Mufti Mahmud\inst{1,2}\orcidID{0000-0002-2037-8348} \and
Jordan J. Bird\inst{1}\orcidID{0000-0002-9858-1231} \and
Thomas Hughes-Roberts\inst{1}\orcidID{0000-0002-3204-8610} \and
David J. Brown\inst{1}\orcidID{0000-0002-1677-7485}}

\institute{Nottingham Trent University, Nottingham, NG11 8NS, United Kingdom \\\email{\{dan.buxton, jordan.bird, thomas.hughes-roberts, david.brown\}@ntu.ac.uk} \and  King Fahd University of Petroleum and Minerals, Dhahran 31261, Kingdom of Saudi Arabia \email{mufti.mahmud@kfupm.edu.sa}}

\authorrunning{D. J. Buxton \textit{et al.}}

\maketitle

\setstretch{1.27}

\begin{abstract}
Digital Human Modelling (DHM) is increasingly shaped by advances in artificial intelligence (AI), wearable biosensing, and interactive digital environments, particularly in research addressing accessibility and inclusion. However, many AI-enabled DHM approaches remain tightly coupled to specific platforms, tasks, or interpretative pipelines, limiting reproducibility, scalability, and ethical reuse. This paper presents a platform-agnostic DHM framework designed to support AI-ready multimodal interaction research by explicitly separating sensing, interaction modelling, and inference readiness. The framework integrates the OpenBCI Galea headset as a unified multimodal sensing layer, providing concurrent Electroencephalogram (EEG), Electromyogram (EMG), Electro-oculogram (EOG), Photoplethysmogram (PPG), and inertial data streams, alongside a reproducible, game-based interaction environment implemented using SuperTux. Rather than embedding AI models or behavioural inference, physiological signals are represented as structured, temporally aligned observables, enabling downstream AI methods to be applied under appropriate ethical approval. Interaction is modelled using computational task primitives and timestamped event markers, supporting consistent alignment across heterogeneous sensors and platforms. Technical verification via author self-instrumentation confirms data integrity, stream continuity, and synchronisation; no human-subjects evaluation or AI inference is reported. Scalability considerations are discussed with respect to data throughput, latency, and extension to additional sensors or interaction modalities. Illustrative use cases demonstrate how the framework can support AI-enabled DHM and HCI studies, including accessibility-oriented interaction design and adaptive systems research, without requiring architectural modifications. The proposed framework provides an emerging-technology-focused infrastructure for future ethics-approved, inclusive DHM research.

\end{abstract}

\section{Introduction}
Digital Human Modelling (DHM) plays a central role in the design of human–computer systems across domains such as ergonomics, safety, health, and accessibility. Recent advances in wearable sensing and interactive technologies have expanded the range of signals available for modelling human interaction, including neurophysiological, muscular, ocular, and cardiovascular measures. At the same time, there is growing recognition that accessibility and inclusion must be treated as first-class design considerations within DHM, particularly when research aims to support diverse populations and contexts.

Despite these advances, many existing digital modelling and multimodal interaction approaches remain tightly coupled to specific platforms, experimental setups, or task environments. Sensing, interaction, and interpretation are often integrated within bespoke pipelines optimised for a single study or application, limiting reproducibility, portability, and ethical reuse. This coupling presents challenges for accessibility-oriented research, where interaction tasks and sensing configurations may need to be adapted to accommodate differing motor, sensory, or cognitive needs without re-engineering the entire system.

In parallel, the use of neurophysiological signals in human–computer interaction has raised important ethical considerations. While such signals can provide valuable contextual information about interaction, their interpretation is frequently conflated with inference about internal cognitive or emotional states. For DHM research, particularly in accessibility-sensitive contexts, there is a need for infrastructures that clearly separate data acquisition from interpretation, allowing physiological and interaction data to be treated as descriptive observables rather than diagnostic indicators.

This paper addresses these challenges by presenting a platform-agnostic multimodal DHM framework that decouples neurophysiological sensing, interaction modelling, and inference readiness through a modular abstraction architecture. The framework integrates the OpenBCI Galea headset as a unified sensing layer, providing concurrent neurophysiological and inertial data streams, alongside a reproducible, game-based interaction environment implemented using SuperTux. Interaction is modelled through structured task primitives and timestamped event markers, enabling consistent alignment between sensing and interaction while remaining independent of specific hardware or software platforms.

The contribution of this work is architectural rather than evaluative. Technical verification is limited to the authors' self-instrumentation to confirm data integrity, stream continuity, and temporal alignment; no human-subjects research is reported, and no behavioural, emotional, or accessibility outcomes are inferred. By focusing on infrastructure rather than inference, the proposed framework provides a reusable scaffold for future ethics-approved DHM studies, supporting inclusive and accessible research design through platform-independent sensing and interaction modelling.

This paper is organised into the following sections:

\textbf{Related Works}: reviews prior research in Digital Human Modelling, multimodal physiological sensing, game-based interaction, accessibility, and ethical considerations, positioning the present work within existing DHM and HCI literature while identifying limitations in portability, abstraction, and ethical separation.

\textbf{Framework Overview}: introduces the design objectives and architectural principles of the proposed platform-agnostic DHM framework, including separation of sensing, interaction modelling, and inference readiness, with emphasis on accessibility-oriented and ethically bounded research design.

\textbf{Sensing Integration and Verification}: describes the integration of the OpenBCI Galea headset as a multimodal sensing layer, detailing signal abstraction, temporal synchronisation, technical verification via author self-instrumentation, and considerations for scalability and data throughput.

\textbf{Interaction Modelling and Applied Implications}: presents the game-based interaction environment and interaction primitives, followed by illustrative DHM and HCI use cases and concrete accessibility adaptation examples that demonstrate how the framework may support inclusive research without embedding evaluative or diagnostic assumptions.

\textbf{Conclusion}: summarises the contribution and limitations of the framework and outlines planned ethics-approved validation steps and future research directions.

\section{Related Works}
Human Modelling has a long history within ergonomics, safety, and human-system interaction, where computational representations of human characteristics are used to inform system design rather than to evaluate individual performance \cite{bubb2009dhm,chaffin2007dhm}. Early DHM research established that modelling need not be limited to visual or biomechanical avatars, but can instead operate at the level of interaction structure and task abstraction \cite{duffy2017dhm}. Layered DHM architectures separating data acquisition, abstraction, and modelling have subsequently been advocated to support reuse across application domains and experimental contexts \cite{duffy2017dhm}.

In parallel, research in physiological computing has demonstrated that signals such as Electroencephalography (EEG), Electromyography (EMG), Electro-oculography (EOG), and cardiovascular measures can be incorporated into interactive systems as additional information channels \cite{fairclough2009physio}. Importantly, foundational work in this area treats physiological signals as interaction-level observables rather than direct indicators of internal cognitive or emotional state. Multimodal sensing approaches are commonly adopted to improve robustness and contextual coverage in wearable and human-centred systems \cite{banaee2013wearable}, although much of the literature focuses on downstream classification or inference, raising methodological and ethical considerations.

Recent advances in wearable biosensing have enabled compact platforms that integrate multiple physiological and inertial modalities into a single device. The Galea headset\cite{openbci2023galea}, for example, provides concurrent EEG, EMG, EOG, photoplethysmography (PPG), and inertial measurement streams intended for research and interactive applications \cite{bernal2022_galea,gupta2023_galea}. Existing work using similar sensing technologies typically embeds these signals within task-specific pipelines, limiting portability and reuse across studies.

Games and interactive simulations have also been widely used as structured environments \cite{haddick2022} for studying human interaction. Digital games offer deterministic mechanics, repeatable task structures, and well-defined event boundaries, making them suitable as controlled interaction substrates \cite{yannakakis2018aigames}. Prior work has combined gameplay with physiological sensing to model affective or experiential states, often focusing on real-time interpretation or performance evaluation \cite{mandryk2007emotion,nacke2008flow}. In contrast, more neutral uses of games treat them as task environments that generate structured interaction events without embedding interpretative assumptions, supporting reproducible modelling approaches.

Accessibility and inclusion have increasingly been framed within HCI as systems-level design challenges rather than properties to be assessed post hoc \cite{stephanidis2019grand}. Inclusive design approaches emphasise flexibility and adaptability at the level of interaction and infrastructure, enabling accommodation of diverse user needs \cite{newell2000inclusive,abascal2016accessibility}. From a DHM perspective, platform-agnostic sensing and interaction pipelines can therefore support inclusive research design by reducing dependence on proprietary tools or rigid experimental protocols.

Finally, the ethical use of physiological data in interactive systems has received growing attention. Concerns regarding over-interpretation, unintended inference, and misuse of biosignals motivate a clear separation between data acquisition and interpretation \cite{nebeker2019ethics}. Ethical frameworks for human-centred AI similarly emphasise transparency and boundary-setting in sensitive application domains \cite{floridi2018ai4people}. These considerations motivate DHM frameworks that prioritise abstraction and infrastructure over inference, enabling future ethics-approved studies without premature or unsupported claims. 

In comparison to existing DHM and multimodal interaction frameworks, which often integrate sensing, task execution, and interpretation within tightly coupled and application-specific pipelines, the present work focuses explicitly on the infrastructural layer that precedes inference. Rather than proposing new behavioural metrics, adaptive algorithms, or representational models, the contribution lies in separating sensing, interaction modelling, and inference readiness. This distinction enables platform-agnostic deployment and ethical reuse across studies, addressing limitations in portability and reproducibility observed in prior approaches.

\section{Framework Overview}
This work proposes a platform-agnostic framework for DHM that separates multimodal sensing, interaction modelling, and inference readiness into distinct architectural layers. The objective is to provide reusable research infrastructure that supports ethically bounded, accessibility-oriented DHM studies across diverse application contexts. Rather than introducing new behavioural metrics or interpretative models, the framework focuses on architectural principles that enable reproducible, adaptable, and ethically defensible human–computer interaction research.

\subsection{Design Objectives}
The framework is guided by four core design objectives. First, platform agnosticism ensures that sensing hardware, interaction environments, and downstream analysis components can be substituted or extended without architectural modification. Second, separation of concerns is enforced by decoupling sensing, interaction modelling, and inference, reducing methodological entanglement and supporting ethical reuse of collected data. Third, accessibility-oriented extensibility is treated as a design constraint, enabling interaction tasks and sensing configurations to be adapted for diverse participant needs without redefining the core pipeline. Finally, ethical separation of inference ensures that physiological and interaction data are treated as descriptive observables, avoiding premature interpretation or diagnostic claims.

\subsection{Architectural Overview}
\begin{figure}
    \centering
    \includegraphics[width=1\linewidth]{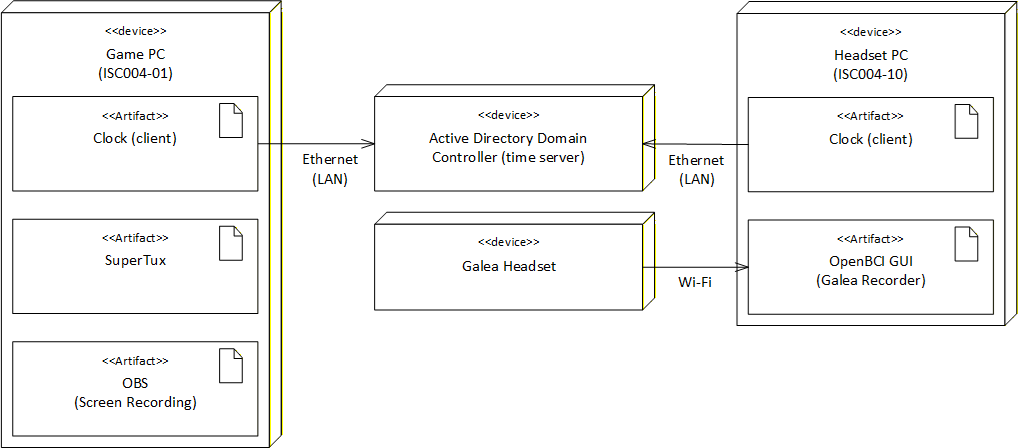}
    \caption{High-level system architecture and deployment of the SuperTux interaction environment and Galea sensing pipeline.}
    \label{fig:tux-galea-deployment}
\end{figure}

At a high level, the framework comprises a multimodal sensing layer, an abstraction layer responsible for temporal alignment and data structuring, and an interaction modelling layer. Physiological and inertial signals are captured independently of the interaction environment and synchronised using timestamped event markers. Interaction is represented through structured task descriptors rather than performance metrics or behavioural scores. This layered architecture supports reuse across DHM applications while maintaining transparency regarding system scope and limitations (Fig. ~\ref{fig:tux-galea-deployment}).

\clearpage

\section{Sensing Integration and Verification}
The multimodal sensing layer integrates the OpenBCI Galea headset as a unified source of physiological data. Galea provides concurrent EEG, EMG, EOG, PPG, and inertial measurement streams, enabling capture of interaction-adjacent signals within a single wearable platform. The framework treats these signals as parallel data sources, abstracted from any task-specific interpretation. Table~\ref{tbl:galea-modalities} shows an overview of the modalities available.

\renewcommand{\arraystretch}{1.4}
\begin{table}[htbp!]
    \centering
    
    \caption{Available Galea Beta headset modalities} \label{tbl:galea-modalities}
    
    \begin{tabular}{c|c|c|c|c}
        \textbf{Modality} & \textbf{Location} & \textbf{Sample Rate} & \textbf{Channels} & \textbf{Parameters and Notes} \\ \hline
        
        EEG & Scalp & 250 Hz & 10 & \makecell{Dry active electrodes,\\F1, F2, C3, Cz, C4, P3,\\ Pz, P4, O1, O2} \\ \hline
        ExG & Forehead & 250 Hz & 0-2 & \makecell{Passive EEG\\Fp1, Fp2} \\ \hline
        
        EMG & Facial & 250 Hz & 4-6 & \makecell{Contains ExG} \\ \hline
        
        EOG & Facial & 250 Hz & 2 & \makecell{4 EMG electrodes} \\ \hline
        
        PPG & Ear clip & 250 Hz & n/a & \makecell{Red \& IR light\\A2 clip placement} \\ \hline
        
        IMU & Forehead & 250 Hz & 6-axis & \makecell{Accelerometer with +/- 4g range\\Gyroscope with +/- 500deg/s} \\ \hline
        
        \makecell{IMU\\(MAG)} & Forehead & 25 Hz & 3-axis & \makecell{Magentometer with +/- 1300uT} \\ \hline
    \end{tabular}
\end{table}

\subsection{Signal Abstraction and Synchronisation}
All sensing streams are timestamped at acquisition and aligned with interaction events generated by the task environment. Synchronisation is performed at the abstraction layer, allowing physiological data to be temporally associated with interaction primitives without embedding assumptions about behavioural meaning. This design supports consistent alignment across heterogeneous data sources while preserving flexibility in downstream analysis.

\subsection{Technical Validation}
Technical verification was conducted exclusively through the authors' self-instrumentation to confirm system functionality, stream continuity, and temporal alignment. Verification focused on validating end-to-end data capture and synchronisation rather than behavioural analysis. No human-subjects research was performed, and no behavioural, emotional, or accessibility outcomes were analysed.

\subsection{Scalability and Performance}
The modular separation of sensing and interaction layers supports scalability to larger studies or additional sensors by treating each data stream as an independent, timestamped source. Buffering and decoupling between acquisition, storage, and downstream processing allow increased data throughput without architectural change. While formal latency benchmarking is beyond the scope of the present work, configurable sampling rates and parallel stream handling enable future deployment in larger-scale or longitudinal DHM studies.

\section{Integration Modelling and Applied Implications}
\subsection{Interaction Modelling Using Game-Based Tasks}
\begin{figure}
    \centering
    \includegraphics[width=\linewidth]{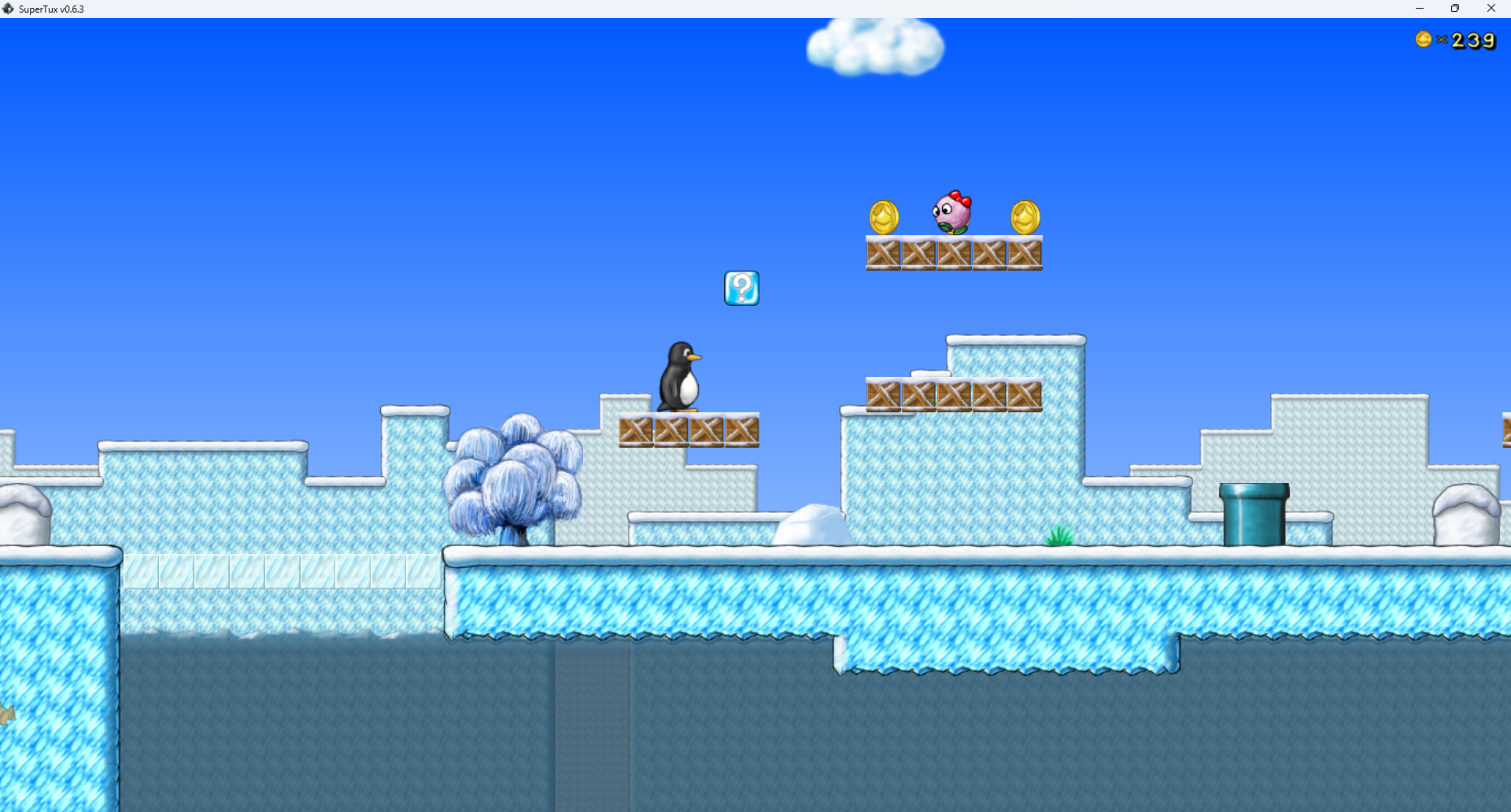}
    \caption{SuperTux game play}
    \label{fig:supertux_screenshot}
\end{figure}

Interaction is implemented using the open-source platform game SuperTux, selected for its deterministic mechanics, discrete event structure, and low sensory complexity. The aim of the game is to reach the end of each level in the shortest amount of time and to gain as many coins as possible, all while avoiding enemy entities that will make the player re-spawn upon contact, in addition to loosing some collected coins and power-up abilities. A screenshot of a level in the game can be seen in Fig~\ref{fig:supertux_screenshot}.

Gameplay actions are abstracted into interaction primitives such as movement sequences, timing events, task progression markers, and error or recovery events. These primitives are independent of both the game engine and sensing hardware, enabling structured modelling of interaction without reliance on game-specific representations.

Interaction descriptors are treated as neutral representations of task engagement rather than indicators of performance quality, cognitive state, or affect. This distinction ensures that interaction modelling remains ethically bounded and compatible with diverse DHM methodologies.

\subsection{Illustrative Modelling and HCI Use Cases}
Although no human-subjects evaluation is reported, the framework is designed to support a range of DHM and HCI research scenarios. For example, future ethics-approved studies could use the interaction and sensing pipeline to examine adaptive interface timing by analysing how physiological and interaction signals co-occur during repeated task exposure. Similarly, the framework could support comparative studies of interaction strategies under different task constraints or input configurations, without modifying the underlying sensing or synchronisation infrastructure. These use cases are illustrative and do not imply evaluation or effectiveness claims.

\subsection{Accessibility and Inclusion Implications}
Accessibility and inclusion are addressed as infrastructural design considerations rather than evaluated outcomes. Interaction tasks can be configured to reduce motor demands by limiting required inputs or adjusting timing constraints, supporting studies involving participants with motor impairments. Sensory load can likewise be modified through visual or auditory simplification, enabling research with participants who experience sensory sensitivities. Such adaptations occur at the interaction layer and do not require changes to the sensing, abstraction, or synchronisation mechanisms, supporting inclusive DHM research design.

\section{Conclusion}
This work presents a framework-level contribution and reports no human-subjects research. Verification was limited to the authors' self-instrumentation to confirm technical functionality. No behavioural, emotional, or accessibility outcomes are inferred.

\subsection{Future Work}
Future work will involve ethics-approved pilot studies to validate the framework in applied DHM contexts. Planned steps include accessibility-focused deployments, comparative task configurations across interaction modalities, and longitudinal studies examining system robustness across repeated sessions. These studies will enable empirical assessment of the framework’s suitability for inclusive DHM research while preserving the ethical separation between sensing, interaction modelling, and inference established in the present work.

The proposed framework provides a reusable, platform-agnostic scaffold for multimodal DHM research that prioritises abstraction, ethical boundary-setting, and accessibility-oriented design. It is intended to support future empirical studies while avoiding premature interpretative claims, aligning with the goals of DHM research within HCII.

\begin{credits}

\subsubsection{\discintname}
The authors have no competing interests to declare that are relevant to the content of this article.
\end{credits}

\bibliographystyle{splncs04}
\bibliography{paper-refs}
\end{document}